\documentclass[nonblindrev]{informs3} 

\usepackage{balance}
\usepackage{amsmath}
\usepackage[tight,footnotesize]{subfigure}
\usepackage{graphicx}
\usepackage{bm}
\usepackage{algorithm}
\usepackage[noend]{algorithmic}
\usepackage{multirow}
\usepackage{slashbox}
\usepackage{caption}
\usepackage{amsfonts}
\usepackage{siunitx}

\OneAndAHalfSpacedXII 

\usepackage{natbib}
 \bibpunct[, ]{(}{)}{,}{a}{}{,}%
 \def\bibfont{\small}%
\TheoremsNumberedThrough     

\EquationsNumberedThrough    

\MANUSCRIPTNO{2015} 

\begin{document}


\RUNAUTHOR{Tang}

\RUNTITLE{Non-monotone Two-Stage Submodular Maximization}

\TITLE{Data Summarization beyond Monotonicity: Non-monotone Two-Stage Submodular Maximization}

\ARTICLEAUTHORS{%
\AUTHOR{Shaojie Tang}
\AFF{Naveen Jindal School of Management, The University of Texas at Dallas}
} 

\ABSTRACT{
The objective of a two-stage submodular maximization problem is to reduce the ground set using provided training functions that are submodular, with the aim of ensuring that optimizing new objective functions over the reduced ground set yields results comparable to those obtained over the original ground set. This problem has applications in various domains including  data summarization. Existing studies often assume the monotonicity of the objective function, whereas our work pioneers the extension of this research to accommodate non-monotone submodular functions. We have introduced the first constant-factor approximation algorithms for this more general case. }


\maketitle

\section{Introduction}
In this paper, we are motivated by the application of data summarization \citep{wei2013using,mirzasoleiman2016fast,wei2015submodularity,lin2011class} and tackle the two-stage submodular optimization problem. In these applications, we are often faced with multiple user-specific submodular functions, which are used to evaluate the value of a set of items. A typical objective is to select a set of $k$ items to maximize each submodular function \citep{krause2014submodular}. While maximizing a single submodular function has been widely explored in the literature, the feasibility of existing solutions diminishes when confronted with a substantial number of submodular functions and items. Consequently, our objective is to reduce the size of the ground set in a manner that minimizes the loss when optimizing a new submodular function over the reduced ground set, as compared to the original ground set.

The problem at hand can be framed as a two-stage submodular maximization problem, as initially introduced in \citep{balkanski2016learning}. While the majority of prior studies in this domain presume that each submodular function exhibits monotone non-decreasing behavior, real-world scenarios often involve objective functions that are non-monotone. These instances include feature selection \citep{das2008algorithms}, profit maximization \citep{tang2021adaptive}, maximum cut \citep{gotovos2015non}, and data summarization \citep{mirzasoleiman2016fast}. A significant contribution presented in our work is the development of the first constant-factor approximation algorithm for the non-monotone two-stage submodular maximization problem, with an approximation ratio of $1/2e$. Remarkably, when the objective function is monotone, our algorithm achieves an improved approximation ratio of $(1-1/e^2)/2$, thereby recovering the result presented in \citep{stan2017probabilistic}.

\subsection{Related Work} The problem of non-monotone submodular maximization has garnered substantial attention in the literature \citep{gharan2011submodular,buchbinder2014submodular,tang2021beyond,tang2021pointwise,tang2022group}. The current state-of-the-art solution for this problem, especially when accounting for a cardinality constraint, is a $0.385$-approximation algorithm \citep{buchbinder2019constrained}. However, it is noteworthy that even though each individual objective function considered in our problem exhibits submodularity, the overall objective function is not submodular in general. As a result, the existing findings on non-monotone submodular maximization do not directly apply to our specific setting.

The most closely related work to our research is the study by \citep{balkanski2016learning,mitrovic2018data} and \citep{stan2017probabilistic}. They have developed constant-factor approximation algorithms, primarily tailored for the monotone case. Our work builds upon and extends their results to address the more general and challenging non-monotone scenario. To achieve this goal, we have integrated the ``local-search'' approach \citep{stan2017probabilistic} with ``sampling'' technique \citep{tang2021beyond} in a non-trivial way, resulting in the creation of a novel sampling-based algorithm. Furthermore, we have incorporated a trimming phase into our algorithm, enabling us to attain the first constant-factor approximation ratio for the non-monotone case.

\section{Problem Formulation}
The input of our problem is a set of $n$ items $\Omega$. There is a group of $m$ non-monotone submodular functions $f_1, \cdots, f_m : 2^\Omega \rightarrow \mathbb{R}_{\geq 0}$. Let $\Delta_i(x, A) = f_i(\{x\} \cup A) - f_i(A)$ denote the marginal gain of adding $x$ to the set $A$ when considering the function $f_i$. Here we say $f_i$ is submodular if and only if $\Delta_i(x, A) \geq \Delta_i(x, A')$ for any two sets $A$ and $A'$ such that $A\subseteq A'\subseteq \Omega$, and any item $x\in \Omega$ such that $x\notin A'$.

Our objective is to compute a reduced ground set $S$ of size $l$, where $l \ll n$, such that it yields good performance across all $m$ functions when the choice is limited to items in $S$. Formally, let
\begin{eqnarray}
F(S) = \sum_{i\in [m]}\max_{A\subseteq S: |A|\leq k} f_i(A)
\end{eqnarray}
where $k$ is the size constraint of a feasible solution. Our goal is to find an optimal solution $O \subseteq \Omega$ that maximizes $F$, i.e.,
\begin{eqnarray}
O \in \argmax_{S\subseteq\Omega: |S|\leq l} F(S).
\end{eqnarray}

It is worth mentioning that the objective function $F(\cdot)$ is typically non-submodular, as observed in \citep{balkanski2016learning}. Consequently, classical algorithms designed for submodular optimization may not provide any approximation guarantees.

\section{Algorithm Design and Analysis}
Before presenting our algorithm, we need some additional notations.  For each $i\in[m]$, we define the gain associated with removing an item $y$ and replacing it with $x$ as $\nabla_i(x, y, A) = f_i(\{x\} \cup A\setminus\{y\}) - f_i(A)$. Then for each $i\in[m]$, we define the largest possible gain brought by $x$,  through local-search, with respect to an existing set $A$   as $\nabla_i(x, A)$. Here the local-search can be realized either by directly adding $x$ to $A$ (while maintaining the cardinality constraint) or by substituting it with an item from $A$. Formally,
\begin{eqnarray}
\nabla_i(x, A)=
\begin{cases}
0 & \mbox{if } x\in A\\
\max\{0, \max_{y\in A} \nabla_i(x, y, A), \Delta_i(x, A)\} & \mbox{if } x\notin A \mbox{ and } |A|< k\\
\max\{0, \max_{y\in A} \nabla_i(x, y, A)\} & \mbox{if } x\notin A \mbox{ and } |A|= k
\end{cases}
\end{eqnarray}

Let $\textsf{Rep}_i(x, A)$ represent the item in $A$ that, when replaced by $x$, maximizes the incremental gain while maintaining feasibility. Formally,
\begin{eqnarray}
\textsf{Rep}_i(x, A)=
\begin{cases}
\emptyset & \mbox{if } \nabla_i(x, A)=0\\
\emptyset & \mbox{if } \nabla_i(x, A)>0 \mbox{ and } |A|<k  \\
                                 &\mbox{ and }\max_{y\in A} \nabla_i(x, y, A)< \Delta_i(x, A)\\
\arg\max_{y\in A} \nabla_i(x, y, A) & \mbox{if }\nabla_i(x, A)>0 \mbox{ and } |A|<k \\
&\mbox{ and }\max_{y\in A} \nabla_i(x, y, A)\geq \Delta_i(x, A)\\
\arg\max_{y\in A} \nabla_i(x, y, A) & \mbox{if } \nabla_i(x, A)>0 \mbox{ and }|A|=k
\end{cases}
\end{eqnarray}

Now we are ready to present the design of our algorithm \textsf{Sampling-Greedy} (Algorithm \ref{alg:LPP0}). Throughout the process, \textsf{Sampling-Greedy}  maintains a solution set denoted as $S$, along with a set of feasible solutions $T_i \subseteq S$ for each function $f_i$ (all of which are initially set to empty). In each iteration, it first computes the top $l$ items $M$ from the extended ground set $\Omega$ based on its combined contribution to each $f_i$, indicated by $\sum_{i=1}^{m} \nabla_i(x, T_i)$. That is,
\begin{eqnarray}
M=\argmax_{A\subseteq\Omega: |A|=l} \sum_{x\in A} \sum_{i=1}^{m} \nabla_i(x, T_i).
\end{eqnarray}

Then it randomly selects one item, say $x^*$, from $M$ and adds $x^*$ to $S$.  \textsf{Sampling-Greedy} then verifies if any of the sets $T_i$ can be improved. This can be achieved by either directly adding $x^*$ (while adhering to the cardinality constraint) or substituting it with an item from $T_i$. For each $i\in[m]$, we update $T_i$ if and only if  $\nabla_i(x^*, T_i)>0$.

Note that there might exist some $i\in [m]$ and $x\in T_i$ such that $f_i(T_i)-f_i(T_i\setminus\{x\})<0$. In other words, certain subsets $T_i$ could contain items that provide negative marginal utility to the set $T_i$.  Consequently, we introduce a ``trimming'' phase (Algorithm \ref{alg:LPP1}) to refine each $T_i$ and ensure that no item contributes negative utility to it. This can be achieved through an iterative process of evaluating the marginal utility of each item within $T_i$ and subsequently removing any items with negative marginal utility. By the submodularity of $f_i$, we can show that after this trimming phase, $T_i$ does not contain any items whose marginal utility if negative. It is also easy to verify that the trimming phase does not decrease the utility of our solution. A formal description of these properties is presented in the following lemma.

\begin{algorithm}[hptb]
\caption{\textsf{Sampling-Greedy}}
\label{alg:LPP0}
\begin{algorithmic}[1]
\STATE $S\leftarrow \emptyset, T_i\leftarrow\emptyset (\forall i\in[m])$
\FOR { $j\in[l]$}
\STATE $M=\argmax_{A\subseteq\Omega: |A|=l} \sum_{x\in A} \sum_{i=1}^{m} \nabla_i(x, T_i).$ \label{line:1}
\STATE randomly pick one item $x^*$ from $M$, $S\leftarrow S\cup\{x^*\}$
\FOR { $i\in[m]$}
\IF{ $\nabla_i(x^*, T_i)>0$}
\STATE $T_i\leftarrow T_i\setminus\textsf{Rep}_i(x^*, T_i)\cup \{x^*\}$
\STATE $T_i \leftarrow \textsf{Trim}(T_i, f_i)$
\ENDIF
\ENDFOR
\ENDFOR
\RETURN $S, T_1, T_2, \cdots, T_m$
\end{algorithmic}
\end{algorithm}

\begin{algorithm}[hptb]
\caption{\textsf{Trim}($B$, $f_i$)}
\label{alg:LPP1}
\begin{algorithmic}[1]
\STATE $A\leftarrow B$
\FOR { $x \in A$}
\IF{ $f_i(A)-f_i(A\setminus\{x\})<0$}
\STATE $A\leftarrow A\setminus \{x\}$
\ENDIF
\ENDFOR
\RETURN $A$
\end{algorithmic}
\end{algorithm}

\begin{lemma}
\label{lem:}
Consider any set of items $B \subseteq \Omega$ and a function $f_i$. Assume $A$ is returned from \textsf{Trim}($B$, $f_i$), we have $f_i(A)\geq f_i(B)$ and for all $x\in A$, we have $f_i(A)-f_i(A\setminus\{x\})\geq 0$.
\end{lemma}
\emph{Proof:} The proof that $f_i(A) \geq f_i(B)$ is straightforward, as it follows from the fact that the trimming phase only eliminates items with a negative marginal contribution. We next prove that for all $x\in A$, we have $f_i(A)-f_i(A\setminus\{x\})\geq 0$. We prove this through contradiction. Suppose there exists an item $y \in A$ such that $f_i(A) - f_i(A\setminus\{y\}) < 0$. Let's denote the solution before considering the inclusion of $y$ as $A'$. In this case, it must hold that $f_i(A') - f_i(A'\setminus\{y\}) \geq 0$, as otherwise, the trimming phase would  eliminate $y$ from the solution. Furthermore, it is straightforward to confirm that $A \subseteq A'$. As a consequence, based on the assumption that $f_i$ is a submodular function, we have $f_i(A) - f_i(A\setminus\{y\}) \geq f_i(A') - f_i(A'\setminus\{y\})$. This, together with $f_i(A')-f_i(A'\setminus\{y\})\geq 0$, implies that $f_i(A)-f_i(A\setminus\{y\})\geq f_i(A')-f_i(A'\setminus\{y\})\geq 0$. This contradicts to the assumption that $f_i(A)-f_i(A\setminus\{y\})< 0$. $\Box$
\subsection{Performance analysis}
First, it is easy to verify that \textsf{Sampling-Greedy} requires $O(l(mkl+mn))$ function evaluations. This is because \textsf{Sampling-Greedy} comprises $l$ iterations, where each iteration involves $mkl$ function evaluations in Line \ref{line:1} of Algorithm \ref{alg:LPP0}, along with an additional $mn$ function evaluations in Algorithm \ref{alg:LPP1}. In the following theorem, we show that the expected utility of our solution is at least a constant-factor approximation of the optimal solution.

\begin{theorem}
\textsf{Sampling-Greedy} returns a random set $S$ of size at most $l$ such that
\begin{eqnarray}
\mathbb{E}_S[F(S)] \geq \frac{1}{2e} F(O)
\end{eqnarray} where $O$ represents the optimal solution.
\end{theorem}

The rest of this section is devoted to proving this theorem. The basic idea behind the proof is to establish a lower bound on the expected marginal utility achieved by adding $x^*$ to set $S$ after each iteration. We demonstrate that this utility increment is substantial enough to guarantee a $1/2e$ approximation ratio. Consider an arbitrary round $t\in[l]$ of \textsf{Sampling-Greedy}, let $S$ and $T_1, \cdots, T_m$ denote the solution obtained at the end of round $t$. By the design of \textsf{Sampling-Greedy}, we randomly pick an item $x^*$ from $M$ and add it to $S$, hence, by the definition of $M$, the expected marginal utility of adding  $x^*$ to $S$ before the ``trimming phase'' is
\begin{eqnarray}
\mathbb{E}_{x^*}[\sum_{i=1}^m \nabla_i(x^*, T_i)]=\frac{1}{l} \max_{A\subseteq\Omega: |A|=l} \sum_{x\in A} \sum_{i=1}^{m} \nabla_i(x, T_i).
\end{eqnarray}

Recall that the trimming phase does not decrease utility. Therefore, the ultimate expected utility increment after each iteration is at least $\mathbb{E}_{x^*}[\sum_{i=1}^m \nabla_i(x^*, T_i)]$. Moreover, because $F$ is a monotone function, it is safe to assume that the size of the optimal solution is $l$, i.e, $|O|=l$. We next provide a lower bound on $\mathbb{E}_{x^*}[\sum_{i=1}^m \nabla_i(x^*, T_i)]$.

Observe that
\begin{eqnarray}
&&\mathbb{E}_{x^*}[\sum_{i=1}^m \nabla_i(x^*, T_i)]=\frac{1}{l} \max_{A\subseteq\Omega: |A|=l} \sum_{x\in A} \sum_{i=1}^{m} \nabla_i(x, T_i)~\nonumber\\
&&\geq \frac{1}{|O|}\sum_{x\in O}\sum_{i=1}^m \nabla_i(x, T_i)=\frac{1}{l}\sum_{x\in O}\sum_{i=1}^m \nabla_i(x, T_i)\label{eq:7}
\end{eqnarray}

Let $O_i \subseteq O$ represent a subset with a maximum size of $k$ items, chosen to maximize $f_i$, i.e., $O_i =\argmax_{A\subseteq O: |A|\leq k} f_i(A)$. Inequality (\ref{eq:7}) implies that
\begin{eqnarray}
\mathbb{E}_{x^*}[\sum_{i=1}^m \nabla_i(x^*, T_i)]\geq \frac{1}{l}\sum_{x\in O}\sum_{i=1}^m \nabla_i(x, T_i) \geq \frac{1}{l}\sum_{i=1}^m \sum_{x\in O_i} \nabla_i(x, T_i). \label{eq:8}
\end{eqnarray}

It is easy to verify that there is a mapping $\pi$ between $O_i$ and $T_i$ such that every item of $O_i\cap T_i$ is mapped to itself, and every item of $O_i\setminus T_i$ is mapped to either the empty set or an item in $T_i\setminus O_i$. We next give a lower bound of $\nabla_i(x, T_i)$.
\begin{lemma}
\label{lem:1}For all $i\in[m]$ and $x\in O_i$, we have
\begin{eqnarray}
\nabla_i(x, T_i)\geq  \Delta_i(x, T_i) -   \Delta_i(\pi(x), T_i\setminus \{\pi(x)\}).
\end{eqnarray}
\end{lemma}
\emph{Proof:} We prove this lemma in three cases. We first consider the case when $x\notin T_i$ and $\pi(x)\neq \emptyset$. In this case, the following chain proves this lemma.
\begin{eqnarray}
\nabla_i(x, T_i)&&\geq f_i(\{x\}\cup T_i\setminus\{\pi(x)\}) - f_i(T_i)\\
&&= \Delta_i(x, T_i) -   \Delta_i(\pi(x), T_i\cup\{x\}\setminus \{\pi(x)\})\\
&&\geq \Delta_i(x, T_i) -   \Delta_i(\pi(x), T_i\setminus \{\pi(x)\})
\end{eqnarray}
where the first inequality is by the definition of $\nabla_i(x, T_i)$ and the second inequality is by the assumption that $f_i$ is a submodular function.

We next consider the case when $x\notin T_i$ and $\pi(x)= \emptyset$. In this case, because $\pi(x)= \emptyset$, i.e., $x$ is not mapped to any item from $T_i$, we have $|T_i|<k$. Hence,
\begin{eqnarray}
\label{eq:1}
\nabla_i(x, T_i) = \max\{0, \max_{y\in T_i} \nabla_i(x, y, T_i), \Delta_i(x, T_i)\} \geq \Delta_i(x, T_i).
\end{eqnarray}

 Moreover,  $\pi(x)= \emptyset$ implies that
 \begin{eqnarray}
 \label{eq:2}
 \Delta_i(\pi(x), T_i\setminus \{\pi(x)\})=0.
\end{eqnarray}

 It follows that
\begin{eqnarray}
\nabla_i(x, T_i)\geq \Delta_i(x, T_i) - 0 = \Delta_i(x, T_i) -   \Delta_i(\pi(x), T_i\setminus \{\pi(x)\}),
\end{eqnarray}
where the inequality is by inequality (\ref{eq:1}) and the equality is by equality (\ref{eq:2}).

At last, we consider the case when  $x\in T_i$. In this case, we have $ \Delta_i(x, T_i)=0$, and $\Delta_i(\pi(x), T_i\setminus \{\pi(x)\})\geq 0$, a consequence of the trimming phase (Lemma \ref{lem:}). Hence, $\Delta_i(x, T_i) -   \Delta_i(\pi(x), T_i\setminus \{\pi(x)\})\leq 0$. It follows that
\begin{eqnarray}
\nabla_i(x, T_i)\geq  0 \geq \Delta_i(x, T_i) -   \Delta_i(\pi(x), T_i\setminus \{\pi(x)\}).
\end{eqnarray} $\Box$

Inequality (\ref{eq:8}) and Lemma \ref{lem:1} imply that
\begin{eqnarray}
\mathbb{E}_{x^*}[\sum_{i=1}^m \nabla_i(x^*, T_i)]&&\geq \frac{1}{l}\sum_{i=1}^m \sum_{x\in O_i} \nabla_i(x, T_i)\\
&&\geq \frac{1}{l}\sum_{i=1}^m \sum_{x\in O_i} (\Delta_i(x, T_i) -   \Delta_i(\pi(x), T_i\setminus \{\pi(x)\})). \label{eq:3}
\end{eqnarray}

Because $f_i$ is submodular, we have
\begin{eqnarray}
\sum_{x\in O_i} \Delta_i(x, T_i) \geq f_i(O_i \cup T_i)-f_i(T_i). \label{eq:4}
\end{eqnarray}

Moreover, no two items from $O_i$ are mapped to the same item from $T_i$, we have
\begin{eqnarray}
\sum_{x\in O_i}  \Delta_i(\pi(x), T_i\setminus \{\pi(x)\}) \leq \sum_{y \in T_i}  \Delta_i(y, T_i\setminus \{y\}) \leq f_i(T_i) \label{eq:5}
\end{eqnarray}
where the first inequality is by the observation that $ \Delta_i(y, T_i\setminus \{y\})\geq 0$ for all $y \in T_i$ and  the second inequality is by the assumption that $f_i$ is submodular.

Inequalities (\ref{eq:3}), (\ref{eq:4}) and (\ref{eq:5}) together imply that
\begin{eqnarray}
\mathbb{E}_{x^*}[\sum_{i=1}^m \nabla_i(x^*, T_i)]&&\geq \frac{1}{l}\sum_{i=1}^m \sum_{x\in O_i} (\Delta_i(x, T_i) -   \Delta_i(\pi(x), T_i\setminus \{\pi(x)\})) \\
&& \geq  \frac{1}{l}\sum_{i=1}^m (f_i(O_i \cup T_i)-f_i(T_i) - f_i(T_i))\\
&&=  \frac{1}{l}\sum_{i=1}^m (f_i(O_i \cup T_i)-2 f_i(T_i)). \label{eq:16}
\end{eqnarray}

Taking the expectation over $T_1, \cdots, T_m$ for both the left and right hand sides of  (\ref{eq:16}), we have
\begin{eqnarray}
&&\mathbb{E}_{T_1,\cdots, T_m}\big[\mathbb{E}_{x^*}[\sum_{i=1}^m \nabla_i(x^*, T_i)]\big]\\
&&\geq \mathbb{E}_{T_1,\cdots, T_m}[ \frac{1}{l}\sum_{i=1}^m (f_i(O_i \cup T_i)-2 f_i(T_i))]\\
&& =  \mathbb{E}_{T_1,\cdots, T_m}[ \frac{1}{l}\sum_{i=1}^m (f_i(O_i \cup T_i))]- \mathbb{E}_{T_1,\cdots, T_m}[\sum_{i=1}^m \frac{2}{l}  f_i(T_i))] \\
&&=\frac{1}{l}  \mathbb{E}_{T_1,\cdots, T_m}[ \sum_{i=1}^m (f_i(O_i \cup T_i))]- \frac{2}{l}  \mathbb{E}_{T_1,\cdots, T_m}[\sum_{i=1}^m f_i(T_i))] \label{eq:9}\\
&&\geq \frac{1}{l} (1-\frac{1}{l})^t \sum_{i=1}^m  f_i(O_i) - \frac{2}{l} \mathbb{E}_{T_1,\cdots, T_m}[f_i(T_i))] \\
&& = \frac{1}{l} (1-\frac{1}{l})^t F(O) - \frac{2}{l} \mathbb{E}_{T_1,\cdots, T_m}[f_i(T_i))]. \label{eq:6}
\end{eqnarray}
The second inequality is by the observation that $ \mathbb{E}_{T_1,\cdots, T_m}[ \sum_{i=1}^m (f_i(O_i \cup T_i))]\geq (1-\frac{1}{l})^t \sum_{i=1}^m  f_i(O_i)$. To prove this inequality, recall that in each round, \textsf{Sampling-Greedy} randomly picks an item from $M$ to be included in $S$. Hence, right before entering round $t$ of \textsf{Sampling-Greedy}, each item $x\in \Omega$ has a probability of at most $p=1-(1-\frac{1}{l})^t$ of being included in $S$ and consequently in $T_i$ for all $i\in [m]$. By Lemma 2.2 of \citep{buchbinder2014submodular}, we have $\mathbb{E}_{T_i}[f_i(O_i \cup T_i)]\geq (1-p) f_i(O_i) = (1-\frac{1}{l})^tf_i(O_i)$ for all $i\in[m]$. It follows that $ \mathbb{E}_{T_1,\cdots, T_m}[ \sum_{i=1}^m (f_i(O_i \cup T_i))]\geq (1-\frac{1}{l})^t \sum_{i=1}^m  f_i(O_i)$.

Let $X_t$ denote the value of $\mathbb{E}_{T_1,\cdots, T_m}\big[\mathbb{E}_{x^*}[\sum_{i=1}^m f_i(T_i)]\big]$ at the end of round $t$. Inequality (\ref{eq:6}) implies that
\begin{eqnarray}
X_{t+1}-X_t \geq  \frac{1}{l} (1-\frac{1}{l})^{t}  F(O) - \frac{2}{l} X_t \\
\Rightarrow 2(X_{t+1}-X_t) \geq  \frac{1}{l} (1-\frac{1}{l})^{t}  F(O) - \frac{2}{l} X_t \\
\Rightarrow 2X_{t+1}-2X_t \geq  \frac{1}{l} (1-\frac{1}{l})^{t}  F(O) - \frac{2}{l} X_t\\
\Rightarrow 2X_{t+1} \geq  \frac{1}{l} (1-\frac{1}{l})^{t}  F(O) + (2- \frac{2}{l}) X_t.
\end{eqnarray}

Based on the above inequality, we next prove through induction that $2 X_t \geq \frac{t}{l}(1-\frac{1}{l})^{t-1} F(O)$. Note that $X_0=0$, meaning that the utility before the start of the algorithm is zero.
The induction step is established in the following manner:
\begin{eqnarray}
&&2X_{t+1} \geq  \frac{1}{l} (1-\frac{1}{l})^{t}  F(O) + (2- \frac{2}{l}) X_t\\
\Rightarrow 2X_{t+1} &&\geq  \frac{1}{l} (1-\frac{1}{l})^{t}  F(O) + (1- \frac{1}{l}) \frac{t}{l}(1-\frac{1}{l})^{t-1} F(O)\\
&&=  \frac{1}{l} (1-\frac{1}{l})^{t}  F(O) + \frac{t}{l}(1-\frac{1}{l})^{t} F(O)\\
&&= \frac{t+1}{l}(1-\frac{1}{l})^{t} F(O).
\end{eqnarray}

It follows that the value of $2 X_l$ is at least $(1-\frac{1}{l})^{l-1} F(O) $, which itself is bounded from below by $(1/e) \cdot F(O)$. Here, $X_l$ represents the expected utility of our algorithm upon completion. Hence, the expected utility of our algorithm is at least $X_l \geq (1/2e)\cdot F(O)$.

\subsection{Enhanced results for monotone case} For the case when $f_i$ is both monotone and submodular, we will demonstrate that the approximation ratio of \textsf{Sampling-Greedy} is improved to $(1-1/e^2)/2$ which recovers the results presented in \citep{stan2017probabilistic}. Observe that if $f_i$ is monotone, we have $f_i(O_i \cup T_i)\geq f_i(O_i)$. Hence, inequality (\ref{eq:9}) implies that

\begin{eqnarray}
&&\mathbb{E}_{T_1,\cdots, T_m}\big[\mathbb{E}_{x^*}[\sum_{i=1}^m \nabla_i(x^*, T_i)]\big]\\
&&\geq \frac{1}{l}  \mathbb{E}_{T_1,\cdots, T_m}[ \sum_{i=1}^m (f_i(O_i \cup T_i))]- \frac{2}{l}  \mathbb{E}_{T_1,\cdots, T_m}[\sum_{i=1}^m f_i(T_i))] \\
&&\geq \frac{1}{l}  \mathbb{E}_{T_1,\cdots, T_m}[ \sum_{i=1}^m (f_i(O_i))]- \frac{2}{l}  \mathbb{E}_{T_1,\cdots, T_m}[\sum_{i=1}^m f_i(T_i))]\\
&&= \frac{1}{l}  \sum_{i=1}^m f_i(O_i)- \frac{2}{l}  \mathbb{E}_{T_1,\cdots, T_m}[\sum_{i=1}^m f_i(T_i))]\\
&&= \frac{1}{l}  F(O)- \frac{2}{l}  \mathbb{E}_{T_1,\cdots, T_m}[\sum_{i=1}^m f_i(T_i))] \label{eq:10}
\end{eqnarray}
where the first equality is because $O_i$ is a fixed set for all $i\in[m]$. Let $X_t$ denote the value of $\mathbb{E}_{T_1,\cdots, T_m}\big[\mathbb{E}_{x^*}[\sum_{i=1}^m \nabla_i(x^*, T_i)]\big]$ at the end of round $t$. Inequality (\ref{eq:10}) implies that
\begin{eqnarray}
X_{t+1}-X_t \geq  \frac{1}{l}  F(O) - \frac{2}{l} X_t.
\end{eqnarray}
Previous research \citep{stan2017probabilistic} has demonstrated that by inductively solving the equation above, we can establish that $X_l \geq ((1-1/e^2)/2)\cdot F(O)$.

\bibliographystyle{ijocv081}
\bibliography{reference}




\end{document}